\documentclass[conference]{IEEEtran}
\usepackage[utf8]{inputenc}
\usepackage[T1]{fontenc}
\usepackage{url}
\usepackage{ifthen}
\usepackage{cite}
\usepackage[cmex10]{amsmath}
\usepackage{stfloats}
\usepackage{graphicx}
\usepackage{epstopdf}
\usepackage{amsfonts}
\usepackage{enumitem}
\usepackage{amssymb}
\usepackage{mathrsfs}
\usepackage{bm}
\usepackage{mdwmath}
\usepackage{subfigure}
\usepackage{mdwtab}
\usepackage{dsfont}
\usepackage{color}
\usepackage{verbatim}
\usepackage{array}
\usepackage{epstopdf}
\usepackage{multirow}
\usepackage{booktabs}
\usepackage{lineno}
\usepackage{amsmath}
\usepackage{makecell}
\usepackage{algorithmic}
\usepackage{amsthm}

\usepackage{flushend}
\usepackage[ruled,vlined,linesnumbered]{algorithm2e}

\DeclareMathOperator*{\argmin}{arg\,min}
\allowdisplaybreaks

\usepackage{titletoc}

\title{Joint Optimization of DNN Inference Delay and Energy under Accuracy Constraints for AR Applications}

\author{Guangjin Pan$^{\dagger}$, Heng Zhang$^{\dagger}$, Shugong Xu$^{\dagger}$, Shunqing Zhang$^{\dagger}$, and Xiaojing Chen$^{\dagger}$\\
$^{\dagger}$
Key laboratory of Specialty Fiber Optics and Optical Access Networks, \\
Shanghai University, Shanghai, 200444, China\\
Email: \{guangjin\_pan, hengzhang, shugong, shunqing, jodiechen\}@shu.edu.cn}

\begin{document}
    \maketitle

% The paper headers
\markboth{Journal of \LaTeX\ Class Files,~Vol.~XX, No.~XX, January~2022}%
{Shell \MakeLowercase{\textit{et al.}}: Bare Demo of IEEEtran.cls for IEEE Transactions on Magnetics Journals}
% The only time the second header will appear is for the odd numbered pages
% after the title page when using the twoside option.
% 
% *** Note that you probably will NOT want to include the author's ***
% *** name in the headers of peer review papers.                   ***
% You can use \ifCLASSOPTIONpeerreview for conditional compilation here if
% you desire.

% If you want to put a publisher's ID mark on the page you can do it like
% this:
%\IEEEpubid{0000--0000/00\$00.00~\copyright~2015 IEEE}
% Remember, if you use this you must call \IEEEpubidadjcol in the second
% column for its text to clear the IEEEpubid mark.

% use for special paper notices
%\IEEEspecialpapernotice{(Invited Paper)}

% for Transactions on Magnetics papers, we must declare the abstract and
% index terms PRIOR to the title within the \IEEEtitleabstractindextext
% IEEEtran command as these need to go into the title area created by
% \maketitle.
% As a general rule, do not put math, special symbols or citations
% in the abstract or keywords.
\IEEEtitleabstractindextext{%
\begin{abstract}
The high computational complexity and high energy consumption of artificial intelligence (AI) algorithms hinder their application in augmented reality (AR) systems. This paper considers the scene of completing video-based AI inference tasks in the mobile edge computing (MEC) system. We use multiply-and-accumulate operations (MACs) for problem analysis and optimize delay and energy consumption under accuracy constraints. To solve this problem, we first assume that offloading policy is known and decouple the problem into two subproblems. After solving these two subproblems, we propose an iterative-based scheduling algorithm to obtain the optimal offloading policy. We also experimentally discuss the relationship between delay, energy consumption, and inference accuracy.
\end{abstract}

% Note that keywords are not normally used for peerreview papers.
}

% make the title area
\maketitle

% To allow for easy dual compilation without having to reenter the
% abstract/keywords data, the \IEEEtitleabstractindextext text will
% not be used in maketitle, but will appear (i.e., to be "transported")
% here as \IEEEdisplaynontitleabstractindextext when the compsoc 
% or transmag modes are not selected <OR> if conference mode is selected 
% - because all conference papers position the abstract like regular
% papers do.
\IEEEdisplaynontitleabstractindextext
% \IEEEdisplaynontitleabstractindextext has no effect when using
% compsoc or transmag under a non-conference mode.

% For peer review papers, you can put extra information on the cover
% page as needed:
% \ifCLASSOPTIONpeerreview
% \begin{center} \bfseries EDICS Category: 3-BBND \end{center}
% \fi
%
% For peerreview papers, this IEEEtran command inserts a page break and
% creates the second title. It will be ignored for other modes.
\IEEEpeerreviewmaketitle

\section{Introduction}
\IEEEPARstart{T}{he} metaverse requires people to interact between the real world and the virtual world. Augmented reality (AR) is an essential technology for this to happen. With artificial intelligence (AI) technology, AR can carry out more profound scene understanding and more immersive interactions.

AI has played an essential role in many fields, such as automatic speech recognition (ASR), natural language processing (NLP), computer vision (CV), and so on. However, the computational complexity of AI algorithms, especially deep neural networks (DNN), is usually very high. It is challenging to complete DNN inference timely and reliably on mobile devices with limited computation and energy capacity. In \cite{DeepSense}, experiments show that a typical single-frame image processing AI inference task takes about 600 ms even with speedup from the mobile GPU. In addition, continuously executing the above inference tasks can only last up to 2.5 hours on commodity devices. The above issues result in only a few AR applications currently using deep learning. To reduce the inference time of DNNs, one way is to perform network pruning on the neural network \cite{pruning1,pruning2,pruning3}. However, it could be destructive to the model if pruning too many channels, and it may not be possible to recover a satisfactory accuracy by fine-tuning \cite{pruning1}.

Edge AI \cite{EI1,EI2,EI3} is another approach to solving these problems. The integration of mobile edge computing (MEC) and AI technology has recently become a promising paradigm for supporting computationally intensive tasks. The edge DNN inference tasks need to consider three Key Performance Indicators (KPIs), i.e., delay, energy consumption, and accuracy. \cite{EI-energy1} and \cite{EI-energy2} optimize inference task selection to minimize the energy consumption in wireless networks. To optimize the inference delay, \cite{EI-Delay1} uses a tandem queueing model to analyze the queueing and processing delays of DL tasks in multiple DNN partitions. At the same time, \cite{EI-accuracy1} and \cite{EI-accuracy2} optimize the accuracy of inference tasks under the constraints of delay and energy consumption. What's more, \cite{EI-Delay-energy1} joint optimizes the service placement, computational and radio resource allocation to minimize the users' total delay and energy consumption.

In this paper, we consider a multi-user MEC system and assume that each device executes the video-based DNN inference task. Each device can be AR glasses, mobile robots, and so on. Different from the offloading problem of still-based (image-based or single-frame-video-based) DNN inference tasks in existing studies \cite{EI-Delay1, EI-accuracy1,EI-accuracy2}, we focus on the video-based DNN inference tasks, which have higher computational complexity and require more computing energy. We model the problem as a multi-objective optimization problem to optimize delay and energy consumption with the constraint of inference accuracy. The main contributions of this paper are summarized as follows,
\begin{itemize}
  \item{\textit{Multi-dimensional target optimization.}} We formulate the video-based offloading problem as a mixed-integer nonlinear programming problem (MINLP). Unlike existing work\cite{EI-Delay-energy1}, we optimize latency and energy consumption under the constraints of DNN inference accuracy. To the best of our knowledge, this is the first work that considers the relationship between delay, energy, and accuracy at the same time. At the same time, we also explore the trade-off relationship between latency, energy consumption and inference accuracy.
  
  \item{\textit{MAC-based analysis model.}} We explored the relationship between the number of input video frames and delay, energy consumption and inference accuracy. We also use multiply-and-accumulate operations (MACs) for refined modeling for the delay and energy consumption. To our knowledge, this is the first work using MACs to analyze the AI inference offloading problem. 
  \item{\textit{Iteration-based scheduling scheme.}} To solve the optimization problem, we decompose the original problem. First, assuming that the offloading decision is given, we respectively solve optimization problems of the device set that completes the inference locally and the device set that is offloaded to the edge server. Then, to obtain the optimal offloading policy, we propose a iterative-based algorithm and solve the original problem through iteration.
\end{itemize}

The rest of this paper is organized as follows. In Section \ref{Sec2}, we introduce system models, including delay, energy, and accuracy models. In Section \ref{Sec3}, we formulate the joint optimization problem and convert the original problem to make it more tractable. Section \ref{Sec4} gives the solution algorithm of the proposed problem. Numerical results and analysis are presented in Section \ref{Sec5}. Finally, the paper is concluded in Section \ref{Sec6}.

\section{System Model} \label{Sec2}

We consider a multi-user MEC system with one base station (BS) and $N$ mobile devices, denoted by the set $\mathcal{N}=\left\{1,2,\dots N\right\}$. Each device has a camera and needs to accomplish DNN inference tasks. Due to the limitation of device computational resources, DNN inference tasks can be placed on local or edge servers. The limited computational resource will lead to longer computing delay and greater power consumption when the inference task is executed locally. However, when the inference task is executed on the edge server, it will bring additional wireless transmission delay.

\begin{figure}[tb]
\centering 
\includegraphics[height=2.0in,width=2.8in]{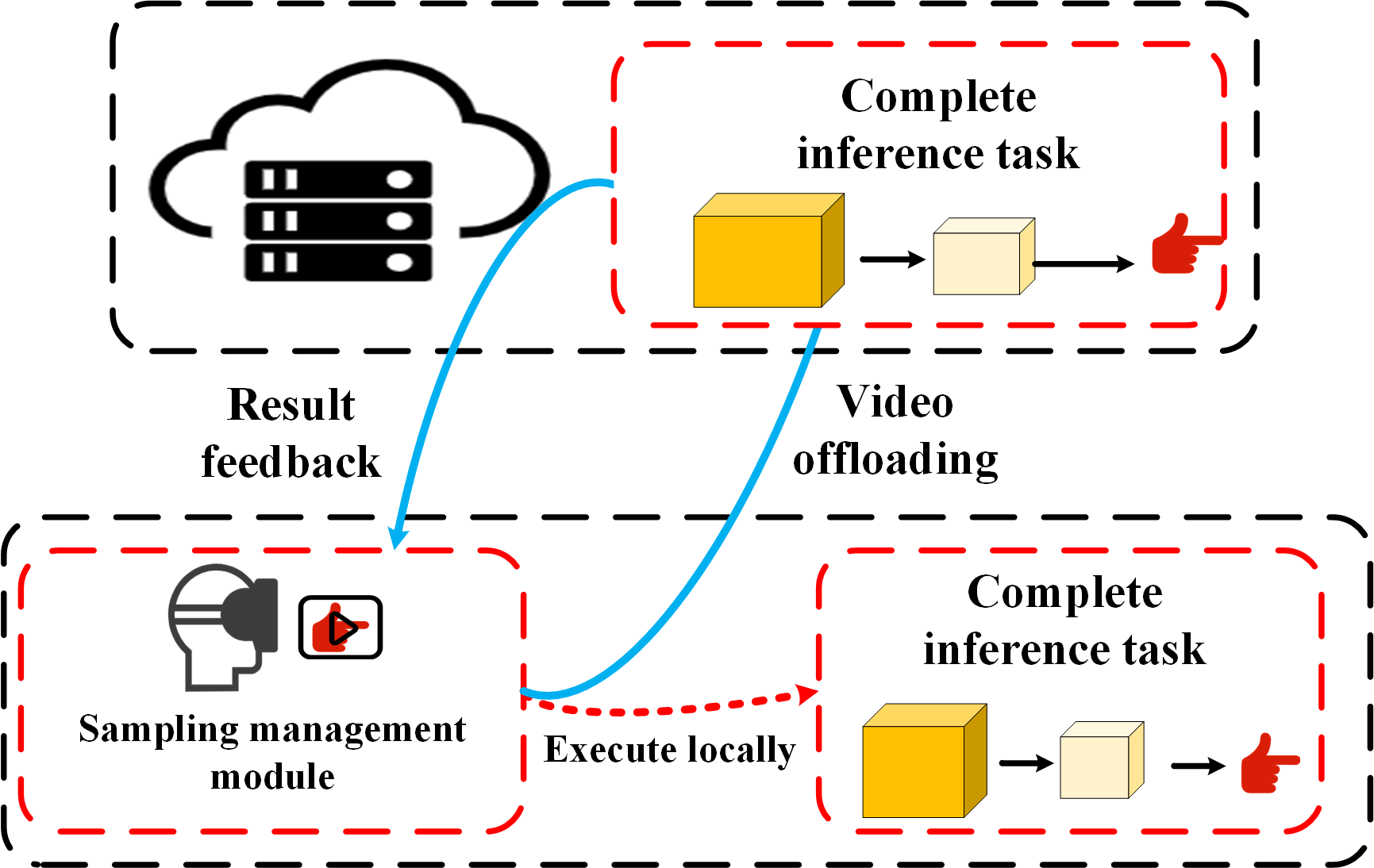} 
\caption{The overview of the video sampling and computing offloading system. The video sampling management module can control the sampling rate of the captured video and determine the number of video frames used for AI inference. Devices can transmit the video to the edge server or perform inference tasks locally based on the wireless channel information and computing capabilities.}
\label{overview} 
\end{figure}

\subsection{Delay and Energy Models for Inference}

The inference delay depends on the DNN model's architecture, the input size of the DNN model, and the device's or server's computing power. The computational complexity of the $n^{th}$ device's task can be expressed as $C(M_{n},\ d_n, \ \Theta_n)$, where $M_{n}$ is the number of input video frames, $d_n$ is the data size of one video frame, and $\Theta_n$ represents the architecture and parameters of the DNN. In this paper, we mainly focus on the impact of the number of input video frames $M_{n}$ on recognition accuracy and the allocation of communication and computing resources. We assume that both $d_n$ and $\Theta_n$ are same for different 
devices. Therefore, we simplify the expression of the computational complexity to $C(M_{n})$. $C(M_{n})$ also means the number of MACs that DNN inference needs to complete, when the number of input video frames is $M_{n}$.

Then we give the expression for the inference delay and energy consumption. The computation delay of the device $n$ and MEC can be respectively expressed as, \begin{eqnarray}
D_n^{md}=\frac{\rho C(M_{n})}{f_n^{md}}, \label{equ2-B-3}\\
D_n^{e}=\frac{\rho C(M_{n})}{f_n^{e}}, \label{equ2-B-4}
\end{eqnarray}
where $\rho$ (cycle/MAC) represents the number of CPU cycles required to complete a multiplication and addition, which depends on the CPU model. $f^{e}_n$ and $f^{md}_n$ (in CPU cycle/s) represent the computing resources allocated by the edge server and the device, respectively. Denote $f^{e,max}$ and $f^{md, max}_n$ to be the total computation resource of the edge server and the mobile device $n$, respectively. Therefore, the computing resources satisfy $\sum_{n \in \mathcal{N}} f_n^{e} \leq f^{e,max}$ and $ f_n^{md} \leq f^{md,max}$.

As for energy consumption, denote $\kappa$ to be a coefficient determined by the corresponding device \cite{EI-accuracy1}. The computational energy consumption of device $n$ can be expressed as,
\begin{eqnarray}
E_n^{md}=\kappa \rho C(M_{n}){f_n^{md}}^2. \label{equ2-B-5}
\end{eqnarray}

\subsection{Delay and Energy Models for Transmission}

We consider a time-division multiple access (TDMA) method for channel access. Denote $R_n$ to be the achievable data rate of device $n$. The delay and energy consumption of transmission can be written as,
\begin{eqnarray}\label{equ2-C-2}
D^t_n=\frac{M_n d_n}{R_nt_n},\\
E^t_n=\frac{M_n d_n}{R_n}p_n.
\end{eqnarray}
where $t_n$ is the proportion of time that device n transmits, and $p_n$ is the transmission power.

\subsection{Inference Tasks Accuracy Model}

\begin{figure}[tb]
\centering 
\includegraphics[height=2.0in,width=2.5in]{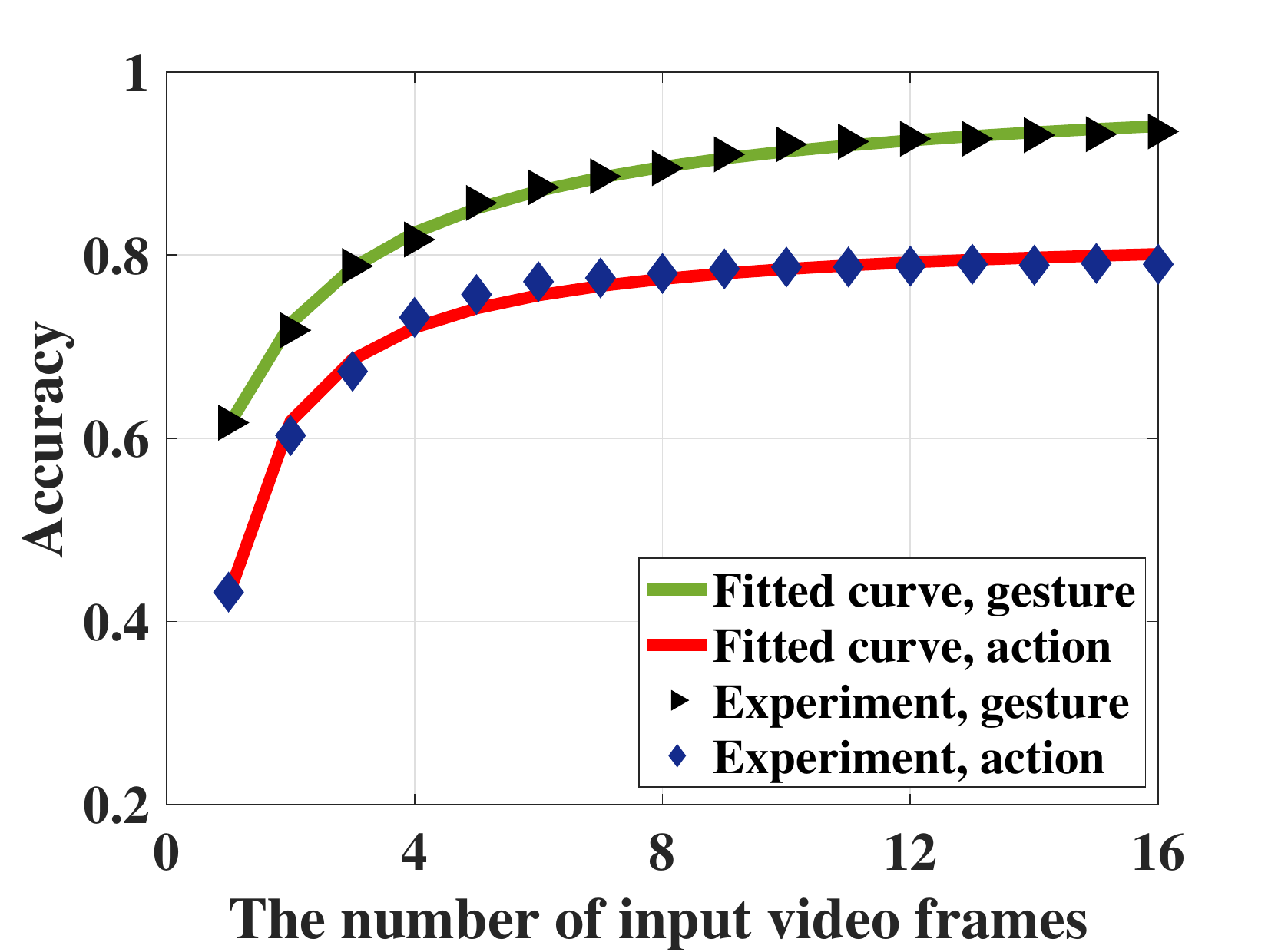} 
\caption{The experimental and fitted curves of gesture recognition task and action recognition task.}
\label{accuracy} 

\vspace{-2 mm}
\end{figure}

As mentioned above, we mainly focus on the impact of the number of input video frames $M_{n}$ on recognition accuracy. We assume that the quality of the input video is the same for different devices. For a certain task and DNN model, the accuracy is only determined by the number of input frames. According to \cite{EI-energy-accuracy1}, more frames will lead to better inference accuracy, and as the input frames continue to increase, the performance gain will gradually decrease. Take gesture and action recognition as examples. Fig. \ref{accuracy} shows the relationship between the inference accuracy and the number of input video frames. Therefore, we define $\Phi({M_{n}})$ as a monotone non-decreasing function to describe the relationship between the accuracy and the number of input frames.

\section{Problem Formulation}  \label{Sec3}
In this section, we formulate the optimization problem to reduce the system's delay and devices' energy consumption under the constraint of recognition accuracy. We analyze the difficulty of solving the problem. To simplify the problem, we make a reasonable conversion of the problem.

\subsection{Original Problem Formulation}
Based on the above analysis, the $n^{th}$ device's delay and energy consumption can be expressed as,
\begin{eqnarray} 
D_n &= & (1-x_n) \frac{\rho C(M_{n})}{f_n^{md}}+x_n(\frac{\rho C(M_{n})}{f_n^{e}}+\frac{ M_{n}d_n}{R_nt_n}),\ \ \ \ \label{equ3-A-1}  \\
E_n  &= &(1-x_n) \kappa \rho C(M_{n}) {f_n^{md}}^2 +x_n(\frac{ M_{n}d_n}{R_n}p_n). \label{equ3-A-2}
\end{eqnarray}
where $x_n$ indicates whether the inference task is executed on local or edge servers.

Given the system model described previously, our goal is to reduce end-to-end delay and energy consumption under the constraint of recognition accuracy. Each device follows the binary offloading policy. The mathematical optimization problem of the total cost can be expressed as,

Problem $\mathcal{P}1$ (\textit{Original Problem}):
\begin{align}
\mathop{\textrm{minimize}}_{\left\{ M_{n}, t_n, f_n^{md} , f_n^e, x_n\right\}}\  & \sum_{n \in \mathcal{N}} \bigg(\beta_1 D_n+\beta_2 E_n\bigg),  \label{equ3-A-3} \\
\textrm{subject to} \ \ \ 
        & \Phi(M_{n}) \ge \alpha_n, \ \forall n \in \mathcal{N}, \tag{\ref{equ3-A-3}a}  \label{equ3-A-3a}\\
       & M_{n} \leq M^{max}_n, \ M_{n} \in \mathbb{Z}, \tag{\ref{equ3-A-3}b}  \label{equ3-A-3b}\\
       & \sum_{n\in \mathcal{N}} x_nt_n \le 1, \tag{\ref{equ3-A-3}c}  \label{equ3-A-3c}\\
       & \sum_{n\in \mathcal{N}}  x_nf_n^e \le f^{e,max} \tag{\ref{equ3-A-3}d},  \label{equ3-A-3d}\\
       & t_n, f_n^e \ge 0, \ \forall n \in \mathcal{N}, \tag{\ref{equ3-A-3}e}  \label{equ3-A-3e}\\
       & 0 \le f_n^{md} \le f^{md, max}_n, \forall n \in \mathcal{N}, \tag{\ref{equ3-A-3}f}  \label{equ3-A-3f}\\
       & x_n \in \left\{0,1\right\}, \forall n \in \mathcal{N},  \tag{\ref{equ3-A-3}g}  \label{equ3-A-3g} 
\end{align}
where $\alpha_n$ represents the recognition accuracy requirement, $\beta_1$, $\beta_2$ are the weight factors. \eqref{equ3-A-3a} represents the recognition accuracy requirement of each device. \eqref{equ3-A-3b} indicates the frame limit for the input video, $\mathbb{Z}$ is the set of integers, and $M^{max}_n$ is the maximum number of frames of the input video. \eqref{equ3-A-3c} and \eqref{equ3-A-3d} represent the communication and computation resource limitation, respectively. \eqref{equ3-A-3f} limits the computation resource of each device.

Problem $\mathcal{P}1$ is a non-convex MINLP problem and is difficult to be solved. First, $C(M_n)$ is an discrete function. As the number of input frames $M_n$ increases, the computational complexity also increases. This kind of increase is irregular because it is affected by the structure of DNN layers, such as the stride and padding size of 3DCNN. Therefore, $C(M_n)$ cannot be used for optimization directly. Second, both $M_n$ and $x_n$ are integers, making the problem difficult to be solved.

\subsection{Problem Conversion}
To make the problem $\mathcal{P}1$ more tractable, we convert the problem. First, we give an approximate expression of the computational complexity function $C(M_n)$, and assume it is a continuous function.

Second, considering that $\Phi(M_{n})$ is a monotone  non-decreasing function. In order not to lose generality, define $M_{n}^{*}=\argmin_{M_n}{\Phi(M_{n})}, \ \Phi(M_{n}) \ge \alpha_n, \ M_{n} \in \mathbb{Z}$. $M_{n}^{*}$ is the minimum number of input frames under the requirement of accuracy. 

We define two sets of devices, i.e. $\mathcal{N}_0=\{n\ | \ x_n=0, n \in \mathcal{N}\}$ and $\mathcal{N}_1=\{n \ | \ x_n=1, n \in \mathcal{N}\}$. $\mathcal{F}_{0,n}$ and $\mathcal{F}_{1,n}$ are the cost function of the device $n$ in sets $\mathcal{N}_0$ and $\mathcal{N}_1$, respectively. As mentioned above, the problem $\mathcal{P}1$ can be rewritten as,

Problem $\mathcal{P}2$ (\textit{Converted Problem}):
\begin{align}
\mathop{\textrm{minimize}}_{\left\{  t_n, f_n^{md} ,f_n^e, x_n\right\}} \ \ &  \sum_{n \in \mathcal{N}_0}(1-x_n)\mathcal{F}_{0,n}({M_n^*},{f_n^{md}}) \nonumber \\ & +\sum_{n \in \mathcal{N}_1}x_n \mathcal{F}_{1,n}({M_n^*},{f_n^e},{t_n}),   \label{equ3-B-3} \\
\textrm{subject to} \  \  
       &\eqref{equ3-A-3c}-\eqref{equ3-A-3g}, \nonumber
\end{align}
where 
\begin{align}
\mathcal{F}_{0,n}({M_n^*},{f_n^{md}}) =  & \   \beta_1\frac{\rho C(M_{n}^*)}{f_n^{md}}+\beta_2 \kappa\rho C(M_n^*)f_n^{md2},\label{equ3-B-4}\\
\mathcal{F}_{1,n}({M_n^*},{f_n^e},{t_n}) = & \  \beta_1 \frac{\rho C(M_{n}^*)}{f_n^{e}}+ \beta_1\frac{M_n^*d_n}{R_nt_n}+\beta_2 \frac{M_n^*d_np_n}{R_n}   \label{equ3-B-5}.
\end{align}

\section{Optimization Problem Solving} \label{Sec4}

In this section, we decompose the problem $\mathcal{P}2$ and propose a iterative-based greedy scheme to solve it. First, supposing that the offloading decision (i.e. $\{x_n\}$) is given, we solve optimization problems of sets $\mathcal{N}_0$ and $\mathcal{N}_1$, respectively. Then, we propose a iterative-based algorithm to optimize the offloading decision $\{x_n\}$.

\subsection{Optimization Problem Solving for $\mathcal{N}_0$}

For set $\mathcal{N}_0$, i.e., when the device executes inference tasks locally, the optimization problem becomes,

Problem $\mathcal{P}_{\mathcal{N}_0}$ (\textit{Problem for $\mathcal{N}_0$}):
\begin{align}
 \mathop{\textrm{minimize}}_{\left\{ f_n^{md} \right\}} \ \  & \mathcal{F}_{\mathcal{P}_{\mathcal{N}_0}} \triangleq  \sum_{n \in \mathcal{N}_0} \mathcal{F}_{0,n}({M_n^*},{f_n^{md}}), \label{equ4-A-1} \\
\textrm{subject to} \ \ \  & \eqref{equ3-A-3f}. \nonumber
\end{align}

The optimization variables in $\mathcal{P}_{\mathcal{N}_0}$ is the local computation resource $f_n^{md}$. We can derive the optimal solution to $\mathcal{P}_{\mathcal{N}_0}$ in a closed-form expression.

\textit{Theorem 1}: The optimal solution to $\mathcal{P}_{\mathcal{N}_0}$ is given by,
\begin{align}
f_n^{md*} =  \ \textrm{min}\{\sqrt[3]{(\frac{\beta_1}{2\beta_2\kappa})}, f^{md, max}_n\} \label{equ4-A-2}
\end{align}

\textit{Proof:}  Please refer to Appendix A.

The partial derivative of $\mathcal{F}_{\mathcal{P}_{\mathcal{N}_0}}$ with respect to $f_n^{md}$ is,
\begin{align}
\frac{\partial \mathcal{F}_{\mathcal{P}_{\mathcal{N}_0}} }{\partial f_n^{md}} = - \beta_1\frac{\rho C(M_{n}^*)}{f_n^{md2}}+2\beta_2 \kappa\rho C(M_n^*)f_n^{md}, \label{equ-appendices-A-1}
\end{align}
By setting $\frac{\partial \mathcal{F}_{\mathcal{P}_{\mathcal{N}_0}} }{\partial f_n^{md}}=0$, we have,
\begin{align}
f_n^{md} = \sqrt[3]{(\frac{\beta_1}{2\beta_2\kappa})}. \label{equ-appendices-A-2}
\end{align} 
Considering the value range of $f_n^{md}$, the optimal solution can be given by,
\begin{align}
f_n^{md*} = & \ \textrm{min}\{\sqrt[3]{(\frac{\beta_1}{2\beta_2\kappa})}, f^{md, max}_n\} \label{equ-appendices-A-3}
\end{align}
which completes the proof.

From \textit{Theorem 1}, we can see that the optimal
local CPU-cycle frequency $f_n^{md}$ is determined by the weight factors $\beta_1$, $\beta_2$, the coefficient of CPU energy consumption $\kappa$, and is limited by its corresponding upper bound $f^{md, max}_n$.

\subsection{Optimization Problem Solving for $\mathcal{N}_1$}
Then we solve the optimization problem of $\mathcal{N}_1$. The problem $\mathcal{P}2$ can be written as,

Problem $\mathcal{P}_{\mathcal{N}_1}$ (\textit{Problem for $\mathcal{N}_1$}):
\begin{align}
 \mathop{\textrm{minimize}}_{\left\{{f_n^e},{t_n} \right\}} \ \ \ & \sum_{n \in \mathcal{N}_1} \mathcal{F}_{1,n}({M_n^*},{f_n^e},{t_n}) ,   \label{equ4-B-1} \\
\textrm{subject to} \ \ \ \ & \eqref{equ3-A-3c}, \ \eqref{equ3-A-3d}, \ \eqref{equ3-A-3e}. \nonumber
\end{align}

The optimization variables in the the problem $\mathcal{P}_{\mathcal{N}_1}$ are the edge computation resource $f_n^e$, and the proportion of transmission time $t_n$. We can obtain the optimal solution to $\mathcal{P}_{\mathcal{N}_1}$ using the method of Lagrange multiplier. The partial Lagrangian function can be written as,
\begin{align}
\mathcal{L}_{\mathcal{P}_{\mathcal{N}_1}} = & \sum_{n \in \mathcal{N}_1} \left( \frac{\beta_1\rho C(M_{n}^*)}{f_n^{e}}+ \frac{\beta_1M_n^*d_n}{R_nt_n}
+\frac{\beta_2 M_n^*d_np_n}{R_n} \right) \nonumber \\ &    +\mu_0 (\sum_{n\in \mathcal{N}_1} t_n - 1) + \mu_1 (\sum_{n\in \mathcal{N}_1} f_n^e - f^{e,max}) , \label{equ4-B-2} 
\end{align}

First of all, according to \eqref{equ4-B-2}, supposing that $M_n^{*}$ is given, we can solve the problem $\mathcal{P}_{\mathcal{N}_1}$ based on the Karush-Kuhn-Tucker (KKT) condition. We can obtain the function expressions of $f_n^{e*}$ and $t_n^{*}$ relative to $M_n^*$, as shown in the following theorem.

\textit{Theorem 2}: The function expressions of $f_n^{e*}$ and $t_n^{*}$ relative to $M_n^{*}$ are given by,
\begin{align}
f_n^{e*} = & \ \frac{f^{e,max} \sqrt{C(M_n^{*})}}{\sum\limits_{i\in \mathcal{N}_1} \sqrt{C(M_i^{*})}}, \label{equ4-B-3}   \\
t_n^{*} = & \ \frac{\sqrt{\frac{M_n^{*}}{R_n}}}{\sum\limits_{i\in \mathcal{N}_1} \sqrt{\frac{M_i^{*}}{R_i}}}. \label{equ4-B-4} 
\end{align}

\textit{Proof:}  Please refer to Appendix B.

According to the KKT conditions, we can obtain the following necessary and sufficient conditions,
\begin{flalign}
& \frac{\partial \mathcal{L}_{\mathcal{P}_{\mathcal{N}_1}} }{\partial f_n^{e*}} = -\frac{\beta_1\rho C(M_{n}^{*})}{f_n^{e*2}}+u_1^{*}=0 ,\ f_n^{e*}>0, \label{equC-1}  \\
& \frac{\partial \mathcal{L}_{\mathcal{P}_{\mathcal{N}_1}} }{\partial t_n^{*}} = -\frac{\beta_1 M_n^{*} d_n}{R_n t_n^{*2}}+u_0^{*}=0,  t_n^{*}>0,\ \label{equC-2}  \\
& \mu_0^{*} (\sum_{n\in \mathcal{N}^{*}} t_n^{*} - 1)=0,  \label{equC-3} \\
& \mu_1^{*} (\sum_{n\in \mathcal{N}^{*}} f_n^{e*} - f^{e,max})=0, \label{equC-4} \\
& \mu_0^{*}, \mu_1^{*} \ge 0. \label{equC-5} 
\end{flalign}

Because $\frac{\beta_1\rho C(M_{n}^{*})}{f_n^{e*2}}$ and $\frac{\beta_1M_n^{*} d_n}{R_n t_n^{*2}}$ are both positive, $\mu_0^{*}$ and $\mu_1^{*}$ are also positive. Therefore, we can obtain,
\begin{flalign}
& \sum_{n\in \mathcal{N}} f_n^{e*} - f^{e,max}=0,  \label{equC-6} \\
& \sum_{n\in \mathcal{N}} t_n^{*} - 1=0, \label{equC-7} \\
& f_n^{e*}=\sqrt{\frac{\beta_1\rho C(M_{n}^{*})}{R_n \mu_1^{*}}}, \label{equC-8} \\
& t_n^{*}=\sqrt{\frac{\beta_1 M_n^{*} d}{R_n\mu_0^{*}}}. \label{equC-9} 
\end{flalign}

Combining \eqref{equC-6} and \eqref{equC-8}, we can get the expression of $f_n^{e*}$ corresponding to $M_n^{*}$,
\begin{align}
f_n^{e*} = & \ \frac{f^{e,max} \sqrt{C(M_n^{*})}}{\sum\limits_{i\in \mathcal{N}_1} \sqrt{C(M_i^{*})}}.
\end{align}
Similarly, combining \eqref{equC-7} and \eqref{equC-9}, we can get the expression of $t_n^{*}$ corresponding to $M_n^{*}$,
\begin{align}
t_n^{*} =  \frac{\sqrt{\frac{M_n^{*}}{R_n}}}{\sum\limits_{i\in \mathcal{N}_1} \sqrt{\frac{M_i^{*}}{R_i}}}.
\end{align}
which completes the proof.

\subsection{Optimization of Offloading Policy $\{x_n\}$}  

We can use the Enumeration method to search for all offloading strategies and get the optimal solution. However, the complexity of Search-based offloading decision algorithm becomes high when the number of devices $N$ grows large. In this section, we propose a iterative-based greedy algorithm to optimize the offloading decision $\{x_n\}$. Then, we analyze the computational complexity of our proposed algorithms.
Inspired by the \textit{Theorem 1} and \textit{Theorem 2}, when executing inference locally, the cost function $\mathcal{F}_{0,n}$ and optimization variables $f_n^{md}$, $M_n$ only depend on the device's own parameters and are not affected by the parameters of other devices. However, for edge set $\mathcal{N}_1$, the cost function is related to the number and parameters of devices in the set $\mathcal{N}_1$. The principle of the iterative-based algorithm is introduced as follows. First, calculate the cost function $\{\mathcal{F}_{0,n}\}$ of set $\mathcal{N}_0$ when each device's task is executed locally. Second, assuming that all devices are offloaded to the edge server for inference and $|\mathcal{N}_1|=N$. In each iteration, the cost function $\{\mathcal{F}_{1,n}\}$ corresponding to each device of $\mathcal{N}_1$ is obtained. Comparing $\{\mathcal{F}_{0,n}\}$ and $\{\mathcal{F}_{1,n}\}$ of the devices in the $\mathcal{N}_1$ set, we can get the difference between $\{\mathcal{F}_{0,n}\}$ and $\{\mathcal{F}_{1,n}\}$ and select the device with the largest difference as $k$. Try to put the device $k$ from the set $\mathcal{N}_1$ into the set $\mathcal{N}_0$ and compute the cost of new sets. If the total cost of new sets is reduced, continue the next iteration. Otherwise, put the device $k$ back to the set $\mathcal{N}_1$.

\section{NUMERICAL RESULTS} \label{Sec5}
In this section, we evaluate the performance of the proposed algorithms via simulations. For all the simulation results, unless specified otherwise, we set the downlink bandwidth as $B_w=5$ MHz and the power spectral as $N_0=-174$ dBm/Hz \cite{EI-accuracy1}. According to \cite{EI-energy1}, the path loss is modelled as $PL=128.1+37.6\log_{10}(D)$ dB, where $D$ is the distance between the device and the BS in kilometres. Devices randomly distributed in the area within  [500m 500m]. The computational resource of the MEC server and devices are set to be 1.8 GHz and 22 GHz, respectively. The recognition accuracy requirement and the maximum number of input video frames are set to $\alpha=0.90, M_n^{max}=16$, respectively. The coefficient $\kappa$ is determined by the corresponding device and is set to be $10^{-28}$ in this paper according to \cite{EI-accuracy1}. The size of the input video is $112 * 112 * M_n$. In addition, the coefficient of computational complexity $\rho$ is set to be 0.12 cycle/MAC, which is obtained through several experiments. Weights $\beta_1, \ \beta_2$ are set to be 0.5, 0.5, respectively. The default number of users is 20. We perform experimental analysis with the gesture recognition task and the Resnet-18 network.

\subsection{Simulation Results of Average Cost}

\begin{figure}[tb]
\centering 
\includegraphics[height=2.2in,width=2.8in]{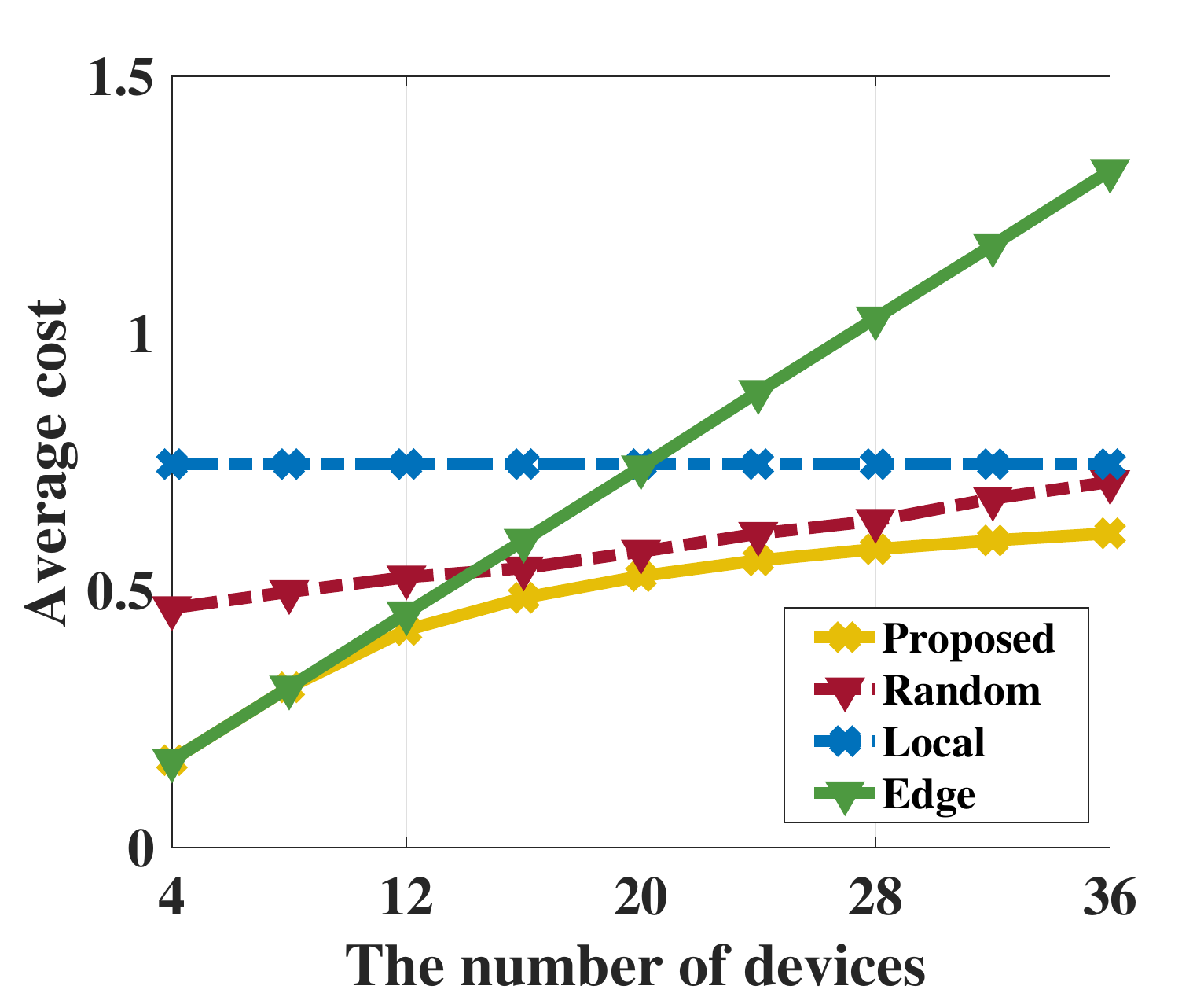} 
\caption{Comparison of the proposed scheme and three baseline schemes. When there are only a small number of devices, the average cost of the proposed scheme and the scheme that devices are randomly offloaded to the edge server to complete inference tasks with 50\% probability.}
\label{result_TC} 
\end{figure}

In this section, we compare proposed algorithms and some baseline algorithms. We mainly compare the average cost. We run 100 tests and can calculate the average cost of each device and the average running time of each test.

\begin{figure*}[t]
\subfigure{\begin{minipage}[t]{0.33\linewidth}
\centering 
\includegraphics[height=1.8in,width=2.2in]{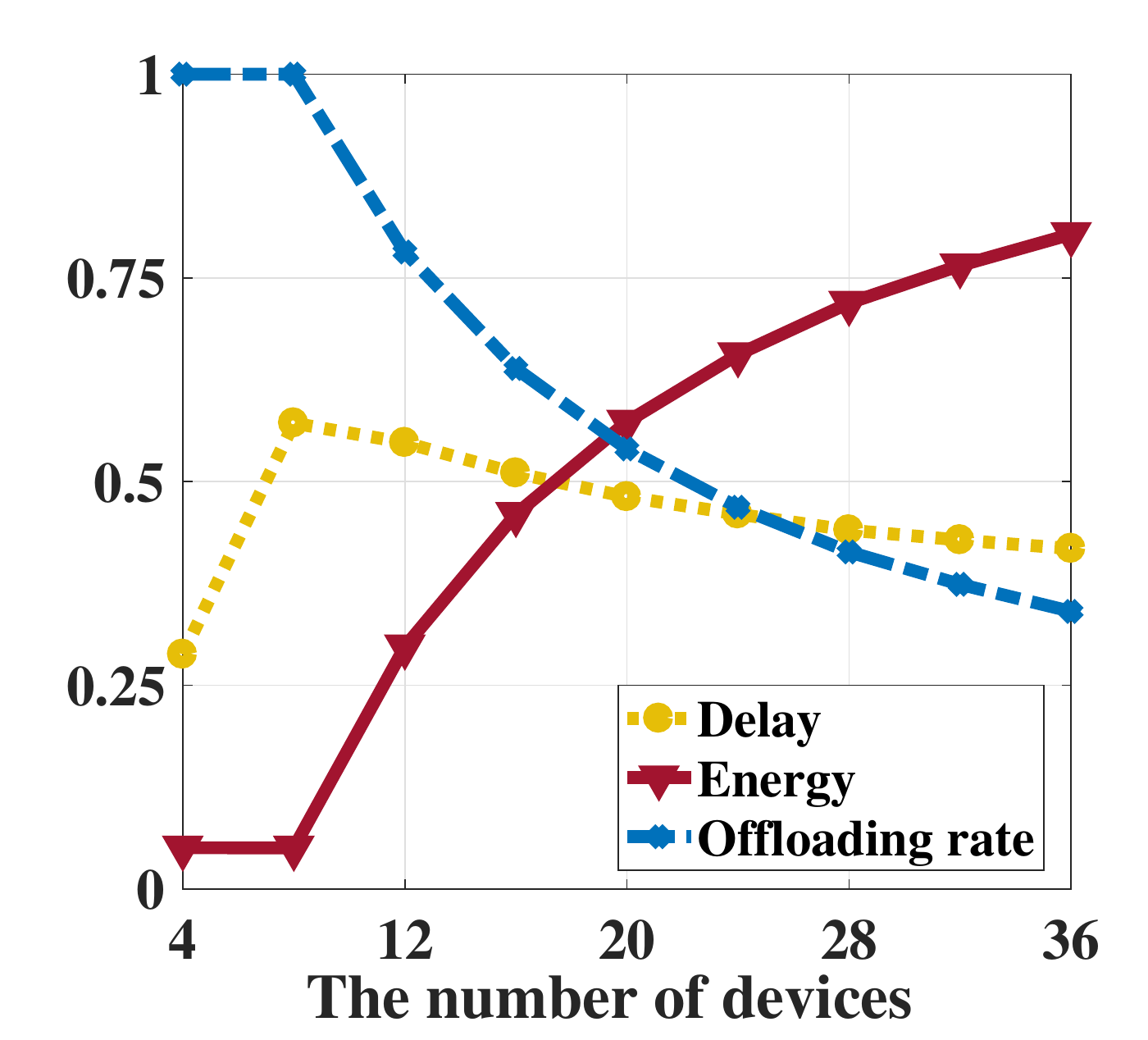} 
\label{result-device} 
\end{minipage}}
\subfigure{
\begin{minipage}[t]{0.33\linewidth}
\centering 
\includegraphics[height=1.8in,width=2.2in]{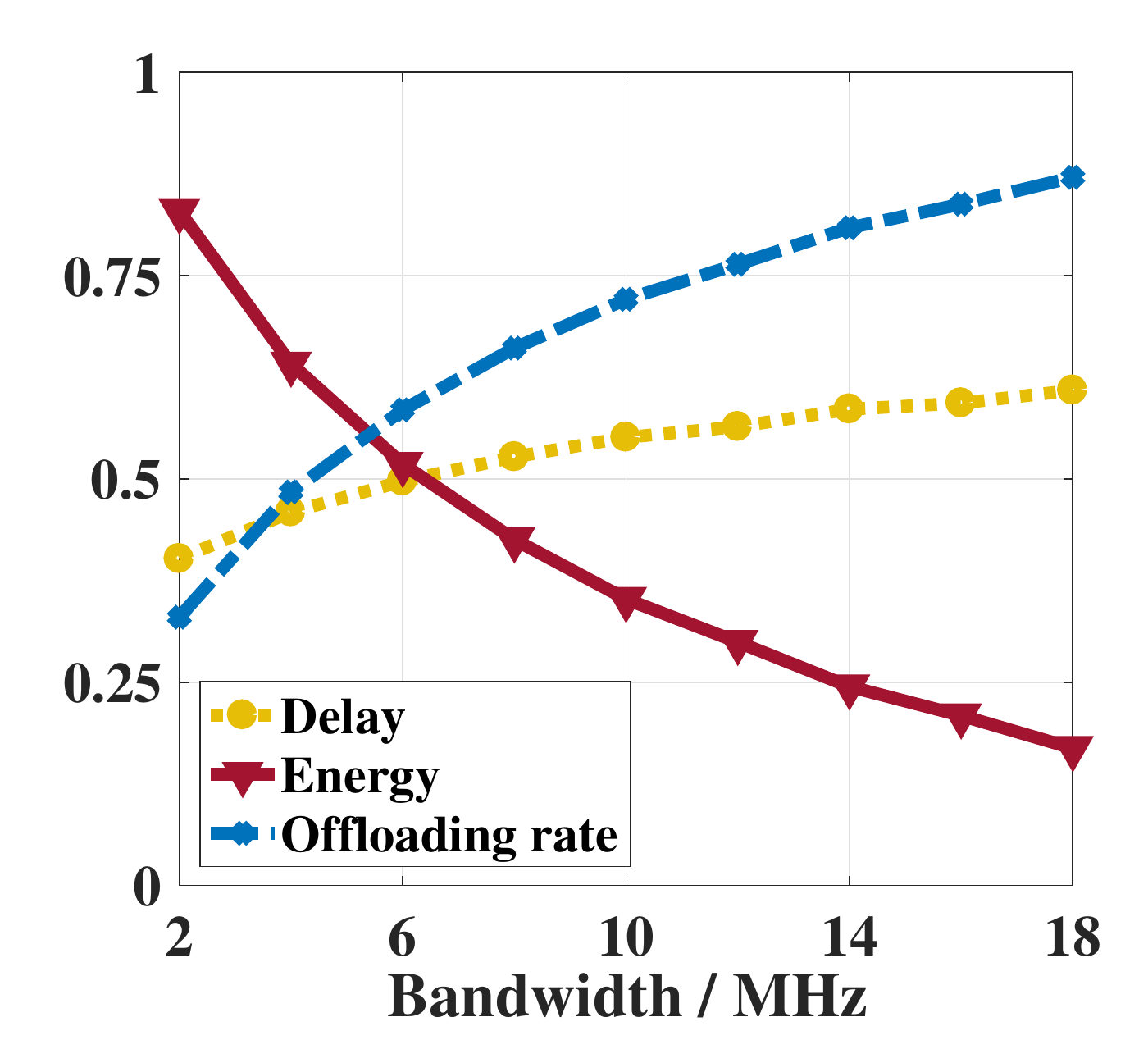} 
\label{result-BW} 
\end{minipage}}
\subfigure{
\begin{minipage}[t]{0.33\linewidth}
\centering 
\includegraphics[height=1.8in,width=2.2in]{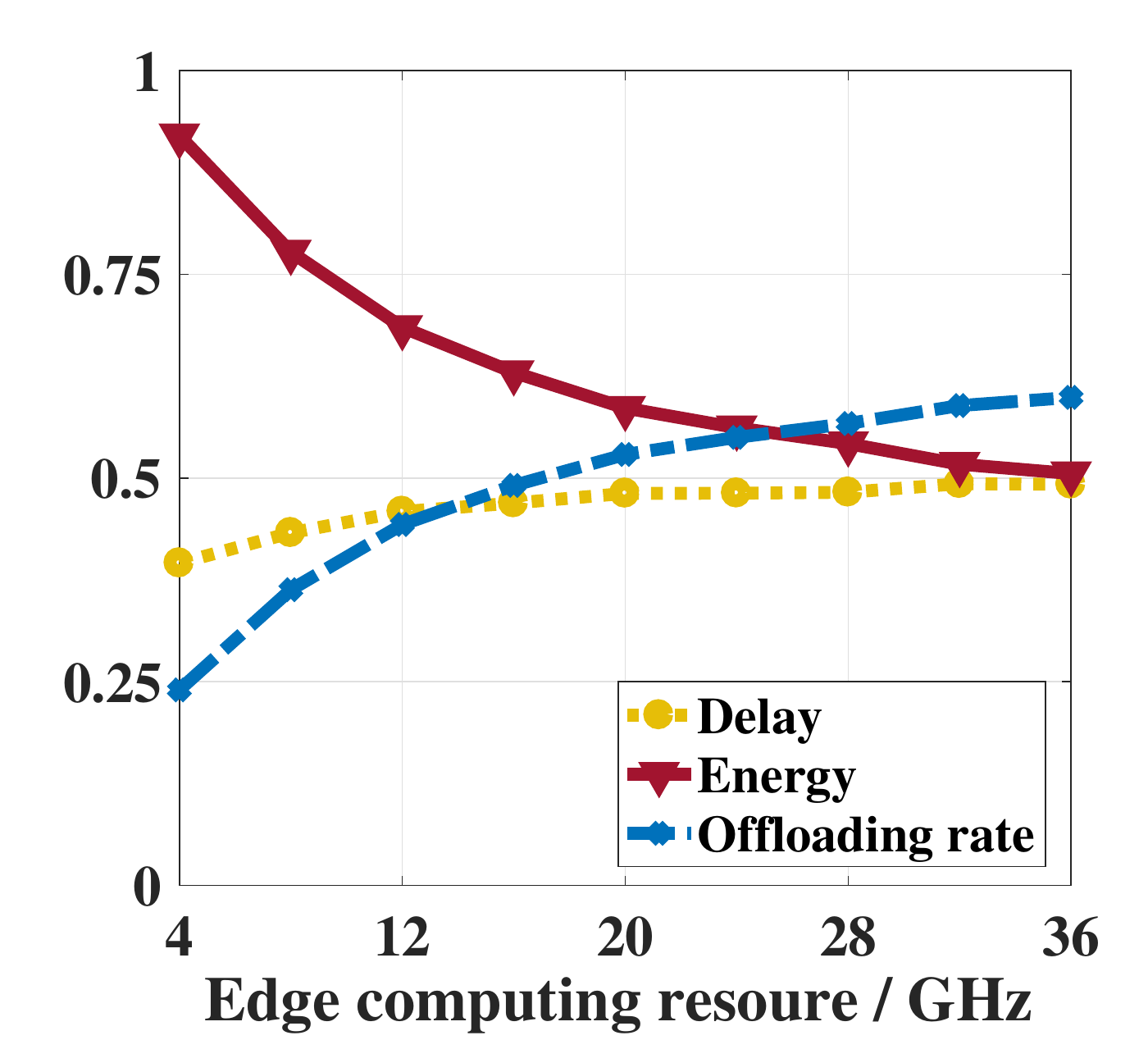} 
\label{result-EC} 
\end{minipage}}
\caption{The average delay, energy, and offloading rate under different numbers of devices, different bandwidths, and different edge computing resources.}
\label{fig:result-C1}
\vspace{-5 mm}
\end{figure*}

In Fig. \ref{result_TC}, we plot the average cost of the proposed scheme and some baseline schemes under a different number of devices. The proposed scheme is compared with the local inference scheme (Local), the edge inference scheme (Edge), and the random offloading scheme (Random). When the number of devices is less than 8, the cost of the scheme that executes tasks only at the edge is almost equal to the cost of the proposed scheme. It is because all devices can benefit from performing inference on the edge server when the number of devices is small. If the inference task is only executed locally, the average cost of the device will not change because the local resources among the equipment do not affect each other. The curve for the Edge scheme is linear because we assume that the AI model is the same for all users.

\subsection{Simulation Results of Delay, Energy, and Accuracy}

This section compares the average delay, energy consumption, and the offloading rate (the proportion of devices that perform inference on the edge server). We set the default requirement for inference accuracy to be 0.9. We consider the different number of devices, different bandwidths, different edge computing resources, and different weights $\beta_1$, $\beta_2$.

Fig. \ref{fig:result-C1} shows the average delay, energy, and offloading rate under different numbers of devices, different bandwidths, and different edge computing resources. In Fig. \ref{result-device}, we plot results with different numbers of devices. As shown in Fig. \ref{result-device}, when the number of devices is small (less than 8), with the number of devices increasing, the average delay increases, and the average energy consumption decrease. This is because when the number of devices is less than 8, all devices offload the task to the edge server (the offloading rate is equal to 1). With the number of devices increasing, communication resources and the edge server's computation resources are shared by more devices. It leads to an increase in the delay and a decrease in the number of input video frames for inference. When the number of devices exceeds 8, the average energy consumption increase, and the average delay and offload rate gradually decrease. Considering different bandwidths and different edge computing resources, we plot Fig. \ref{result-BW} and Fig. \ref{result-EC}. As shown in Fig. \ref{result-BW} and Fig. \ref{result-EC}, the offloading rate increases with the bandwidth and edge computing resource increasing, indicating that more devices are offloading tasks to the edge server.

\begin{figure}[tb]
\centering 
\includegraphics[height=2.2in,width=2.6in]{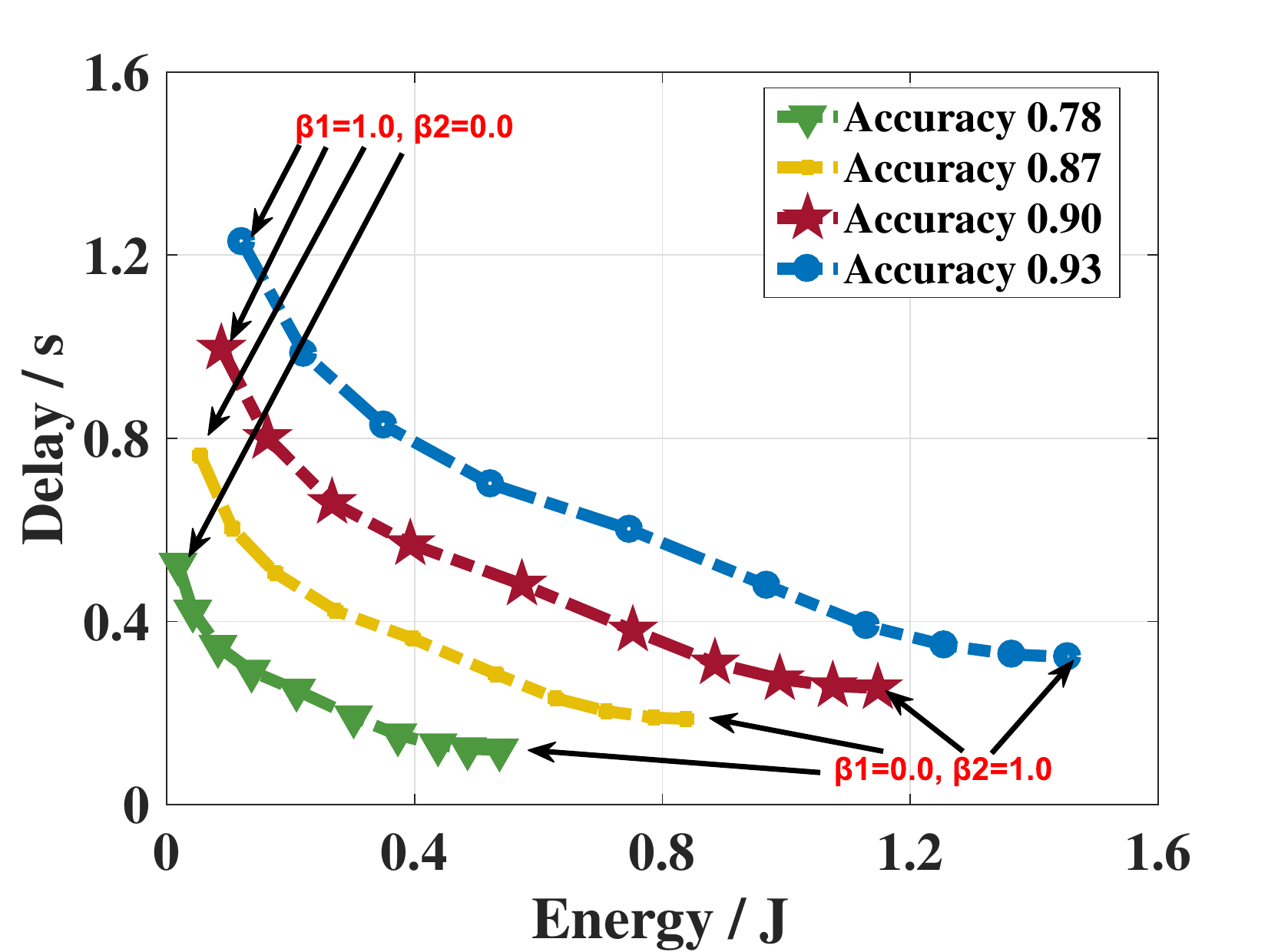}
\caption{The relationship between the delay, energy consumption, and accuracy.}
\label{fig-result-C2}
\end{figure}

We use different weights, $\beta_1, \ \beta_2$ to study the trade-off relationship between the average delay and energy consumption with accuracy requirement. The constraint is $\beta_1+ \beta_2=1$. Fig. \ref{fig-result-C2} shows the delay, energy consumption, and accuracy are mutually limited. Lower energy consumption leads to higher delay when the accuracy is constant. From another perspective, to improve the accuracy, it is necessary to sacrifice the performance of delay and energy consumption.

\section{Conclusion} \label{Sec6}
This paper considers optimizing video-based AI inference tasks in a multi-user MEC system. An MINLP is formulated to minimize the total delay and energy consumption, with the requirement of accuracy. We use a MAC-based model to analyze the problem. We propose an iterative-based scheduling method to solve this problem. We analyze the experimental results with the different number of devices, different bandwidths and different edge computing capabilities. We also plot the relationship curve of the average delay, energy consumption, and accuracy.

\section*{ACKNOWLEDGEMENT}
This work was supported in part by the National Natural Science Foundation of China (NSFC) under Grant 61871262, 62071284, and 61901251,  the National Key R\&D Program of China grants 2017YFE0121400 and 2019YFE0196600, the Innovation Program of Shanghai Municipal Science and Technology Commission grants 20JC1416400 and 21ZR1422400, Pudong New Area Science \& Technology Development Fund, Key-Area Research and Development Program of Guangdong Province grant 2020B0101130012, Foshan Science and Technology Innovation Team Project grant FS0AA-KJ919-4402-0060, and research funds from Shanghai Institute for Advanced Communication and Data Science (SICS).

% if have a single appendix:
%\appendix[Proof of the Zonklar Equations]
% or
%\appendix  % for no appendix heading
% do not use \section anymore after \appendix, only \section*
% is possibly needed

% use appendices with more than one appendix
% then use \section to start each appendix
% you must declare a \section before using any
% \subsection or using \label (\appendices by itself
% starts a section numbered zero.)
%

% use section* for acknowledgment

\bibliographystyle{IEEEtran}

\bibliography{reference}

\end{document}